\documentclass{PHYEAUTH}
\usepackage{graphicx}
\usepackage{amsmath}
\usepackage{amssymb}

\begin{document}

\begin{frontmatter}

\title{Localization in the quantum Hall regime}
\author[address1]{Bernhard Kramer\thanksref{thank1}}, 
\author[address1]{Stefan Kettemann}
and
\author[address2]{Tomi Ohtsuki}

\address[address1]{ I. Institut f\"ur Theoretische Physik, Universit\" at
  Hamburg, Jungiusstra\ss{}e 9, 20355 Hamburg, Germany }

\address[address2]{Department of Physics, 
Sophia University, Kioicho 7-1, Tokyo 102-8554, Japan}

\thanks[thank1]{
Corresponding author. 
E-mail: kramer@physnet.uni-hamburg.de}

\begin{abstract}
  The localization properties of electron states in the quantum Hall regime
  are reviewed. The random Landau model, the random matrix model, the
  tight-binding Peierls model, and the network model of Chalker and Coddington
  are introduced.  Descriptions in terms of equivalent tight-binding
  Hamiltonians, and the 2D Dirac model, are outlined.  Evidences for the
  universal critical behavior of the localization length are summarized. A
  short review of the supersymmetric critical field theory is provided. The
  interplay between edge states and bulk localization properties is
  investigated. For a system with finite width and with short-range
  randomness, a sudden breakdown of the two-point conductance from $ne^{2}/h$
  to 0 ($n$ integer) is predicted if the localization length exceeds the
  distance between the edges.
\end{abstract}

\begin{keyword}
quantum Hall effect \sep localization \sep network model \sep conductance
distribution \sep Anderson transition
\PACS 74.40.Xy \sep 71.63.Hk
\end{keyword}
\end{frontmatter}

\section{Introduction}
The scaling theory of localization \cite{aalr79,mk93} predicts that there is
no Anderson transition (AT) in two dimensions (2D) without interactions. There
are only two known exceptions: One is the quantum Hall transition (QHT), and
the other is the localization-delocalization transition in systems with
spin-orbit interaction that are called the symplectic class. The former has
attracted broad attention after the discovery of the quantum Hall effect
(QHE) by Klaus von Klitzing in 1980 \cite{kvk80}. The symplectic class has
been a subject of continuous effort with a strong increase after the
discovery of the 2D metal-insulator transition (MIT) in Si-MOS-systems
\cite{kravchenko95} which also renewed the interest in whether or
not Coulomb interaction can introduce an MIT in 2D. 

In this paper, we mainly review results for the QHT. It has been noted
immediately after its discovery that the QHE cannot be explained within the
semiclassical Drude model which gives for the longitudinal conductivity
$\sigma_{xx}=0$ in the limit of large magnetic field, while the Hall
conductivity $\sigma_{xy}=\nu e^{2}/h$. If the filling factor $\nu\equiv\rho
h/eB$ is integer ($\rho$ electron density), $\sigma_{xy}$ is quantized in
units of $e^{2}/h$. For explaining the QHE, a certain amount of disorder is
necessary in order to understand why broad plateaus in the Hall conductivity
can be formed and simultaneously the longitudinal conductivity drops to zero:
localized states in the tails of the disorder broadened Landau bands can pin
the Fermi level, but do not contribute to the transport. However, since there
is a non-vanishing Hall conductivity, not all of the states can be localized.
Near the centers of the Landau bands, regions of extended states must exist
that can carry the Hall current \cite{aa81}. That this current has the correct
magnitude for giving integer values of the Hall plateaus first has been
suggested by Prange \cite{p81} who showed that the amount of current which is
lost when a localized state is formed in the tail of a Landau band is exactly
compensated for by the remaining extended states.  Later this has been argued
to be related to a gauge property which was used to conjecture that if an
equilibrium current is flowing in a quantum Hall system this can only give
integer values of the Hall conductance \cite{l81,h82}. For such an equilibrium
current to flow, at least one extended state must exist. Thus the
localization-delocalization phenomenon is closely related to the very
existence of the Hall plateaus.

When introducing the standard models we concentrate predominantly on the
aspect of universality of the transition. We argue that Coulomb interaction
do not change the universality class of the QHT, apart from spin-orbit
interaction which is still a subject of intense ongoing research. We mention
in passing that the scaling properties of the QHT are also the subject of
considerable experimental efforts. So far, temperature and frequency scalings
of the conductances in the region of the QHE are fully consistent with the
picture developed by the theory
\cite{wetal88,ketal91,eetal93,ketal0y,f98,hetal02}.  Furthermore, starting from
the standard Chalker-Coddington network model we report on a class of
equivalent Hamiltonians such as the Dirac model in which the QHT is associated
with a crossover from zero to a finite non-zero mass.

Finally,  recent results for the interplay between localization of bulk
and edge states in systems of finite width are described. It is shown for
short range impurities that if the localization length becomes equal to the
system width, there can be an abrupt breakdown of the two-point quantum Hall
conductance from the plateau value to zero. This is driven by a  dimensional
crossover between 2D and 1D localization of the bulk states 
  resulting in an    abrupt 
chiral metal to insulator 
transition of the edge states. It  resembles
 a first order quantum  phase transition.

\section{The standard models}

\subsection{The random Landau model}
The most straightforward description of the quantum Hall phenomena is obtained
by adding a random potential $V({\bf r})$ to the Hamiltonian of a free
electron moving in a plane in the presence of a perpendicular magnetic field
$B$,
\begin{equation}
  \label{eq:01}
  H=H_{0}+V:=
\frac{1}{2m^{*}}\left({\bf p}+e{\bf A}\right)^{2}+V({\bf r})\,,
\end{equation}
and solving the corresponding Schr\"odinger equation. In the Landau gauge, the
vector potential is ${\bf A}=B(-y,x,0)$; $m^{*}$ and $e$ are the effective
mass of the electron and the elementary charge, respectively, and ${\bf p}$
the momentum operator. The randomness is incorporated by assuming a
distribution function for the potential. In the presence of spatial
correlations,
\begin{equation}
  \label{eq:01a}
 \langle V({\bf r})V({\bf r'})\rangle=W^{2}f({\bf
  r}-{\bf r'})
\end{equation}
where $\langle\ldots\rangle$ denotes the configurational average. For
simplicity, one often also assumes a symmetric potential distribution such
that $\langle V({\bf r})\rangle =0$. If the potential is white noise, $f({\bf
  x})=\delta({\bf x})$. The correlator $f({\bf x})$ reflects the nature and
the range of the impurity potential.

In the representation of the Landau states $|nX\rangle$, $H_{0}|nX\rangle=
E_{n}|nX\rangle$, the above Hamiltonian is
\begin{equation}
  \label{eq:02}
  H=\sum_{nX}E_{n}|nX\rangle\langle nX|+\sum_{nX,n'X'}V_{nX,n'X'}
|nX\rangle\langle n'X'|
\end{equation}
with the matrix elements of the random potential, $V_{nX,n'X'}$, and the
Landau energies $E_{n}=\hbar\omega_{B}(n+1/2)$ ($n=0,1,2\ldots$) that are
degenerate with respect to the quantum number $X\equiv 2\pi
m\ell_{B}^{2}/L_{y}$ (''center-of-motion coordinate'', $m$ integer) with a
degree of $n_{B}=1/2\pi\ell_{B}^{2}$ per unit area. Here,
$\omega_{B}=eB/m^{*}$ and $2\pi\ell_{B}^{2}=\hbar/eB$ are the cyclotron
frequency and the cyclotron area, respectively. The degeneracy corresponds to
the density of the flux quanta $h/e$ in the system. For sufficiently large $B$
the disorder induced coupling between the Landau levels can be neglected.

In this limit, the density of states $\rho(E)$ of a single band can be
obtained exactly \cite{w83,bgi84}. It is symmetric around the Landau band
center and has Gaussian tails for $|E-E_{n}|\gg\Gamma$. Near the band center,
one obtains $\rho(E)=\rho_{n}[1-(E-E_{n})^{2}/\Gamma^{2}]$. For the lowest
band, $\rho_{0}=n_{B}/\pi\Gamma$ and the band width $\Gamma = 2Wn_{B}^{1/2}$.

\begin{figure}[htbp]
\vspace{3mm}\begin{center}
\includegraphics[width=0.8\linewidth]{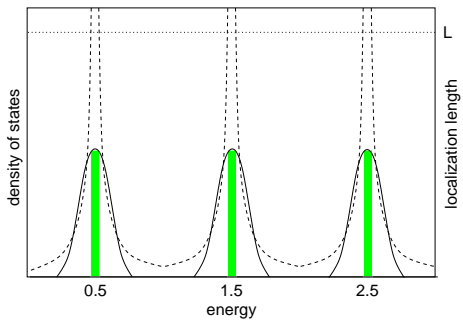}
\end{center}
  \caption{Qualitative picture of the 
    density of states (solid line) and the localization length (dashed) of the
    random Landau model (arbitrary units) as a function of the energy (units
    $\hbar \omega_{B}$). In the energy regions where the localization length
    exceeds a cutoff length $L$ the wave functions are effectively
    delocalized (green regions of the density of states).}
  \label{kvkfig:1}
\end{figure}

Using the Lifshitz argument \cite{l65}, it can be argued that the wave
functions are localized within accidentally formed potentials wells in the
band tails. Towards the band centers, the localization length increases. Using
the above Hamiltonian and with a random superposition of impurity potentials
for modeling the randomness, first quantitative results for the localization
in the lowest Landau bands have been obtained \cite{aa85}. It was found that
the exponential localization length $\xi$ diverges near the centers of the
Landau levels according to power laws
\begin{equation}
  \label{eq:03}
  \xi(E)=\xi_{n}|E-E_{n}|^{-\nu_{n}}\,.
\end{equation}
In the lowest Landau band, an exponent $\nu_{0}\approx 2$ has been found.
Figure \ref{kvkfig:1} shows qualitatively the behavior of the density of
states and the localization length.

\subsection{The random matrix model}
Instead of calculating the matrix elements of the random potential from a
superposition of impurity potentials, one can equivalently start from
(\ref{eq:02}) and choosing the matrix elements of the potential according to
their statistical properties \cite{hk89}. If one is interested only in
universal features the precise form of the matrix elements should not make any
difference. The resulting Hamiltonian corresponds to a random matrix model.
However, in contrast to the conventional random matrix Hamiltonians, the one
derived from the Landau Hamiltonian (\ref{eq:01}) has strongly correlated
matrix elements. For the lowest Landau band and a Gaussian correlator
(\ref{eq:01a}) with correlation length $\sqrt{2}\sigma$
\begin{eqnarray}
  \label{eq:4}
&\!\!\!\!\!\!\!\!\!\!\!\!\!\!\!\!\!\!
\langle V_{XX'}V_{X''X'''}\rangle = \delta_{X-X',X'''-X''}\,
\frac{W^{2}\ell_{B}}{\sqrt{2\pi}\beta L_{\infty}}\nonumber\\
&\nonumber\\
&\quad\times\exp{\left\{-\frac{1}{2}\left[\beta^{2}(X-X')^{2}+
\frac{1}{\beta^{2}}(X-X''')^{2}\right]\right\}}\,
\end{eqnarray}
with the parameter $\beta^{2}:=1+(\sigma/\ell_{B})^{2}$ representing the
correlation of the potential matrix elements along diagonals $X-X'=const$ and
$L_{\infty}$ the size of the system. Even if the starting point is completely
uncorrelated white noise, $\sigma =0$, the correlations in the matrix elements
persist, $\beta =1$. 

From numerical evidence, one can conclude that these correlations are
responsible for the singular behavior of the localization length in the
centers of the Landau bands \cite{jp00}. Mathematically, the properties of
this class of correlated random matrix models have not yet been explored. This
might provide valuable insights into the statistics of spectral properties
like the level statistics and their critical behavior. Eventually, such an
investigation of the properties of correlated banded matrices also might give
insight into the generic mathematical structure behind the phenomenon of the
QHT.

\begin{figure}[htbp]
\hspace{0.45cm}
\includegraphics[width=0.7\linewidth]{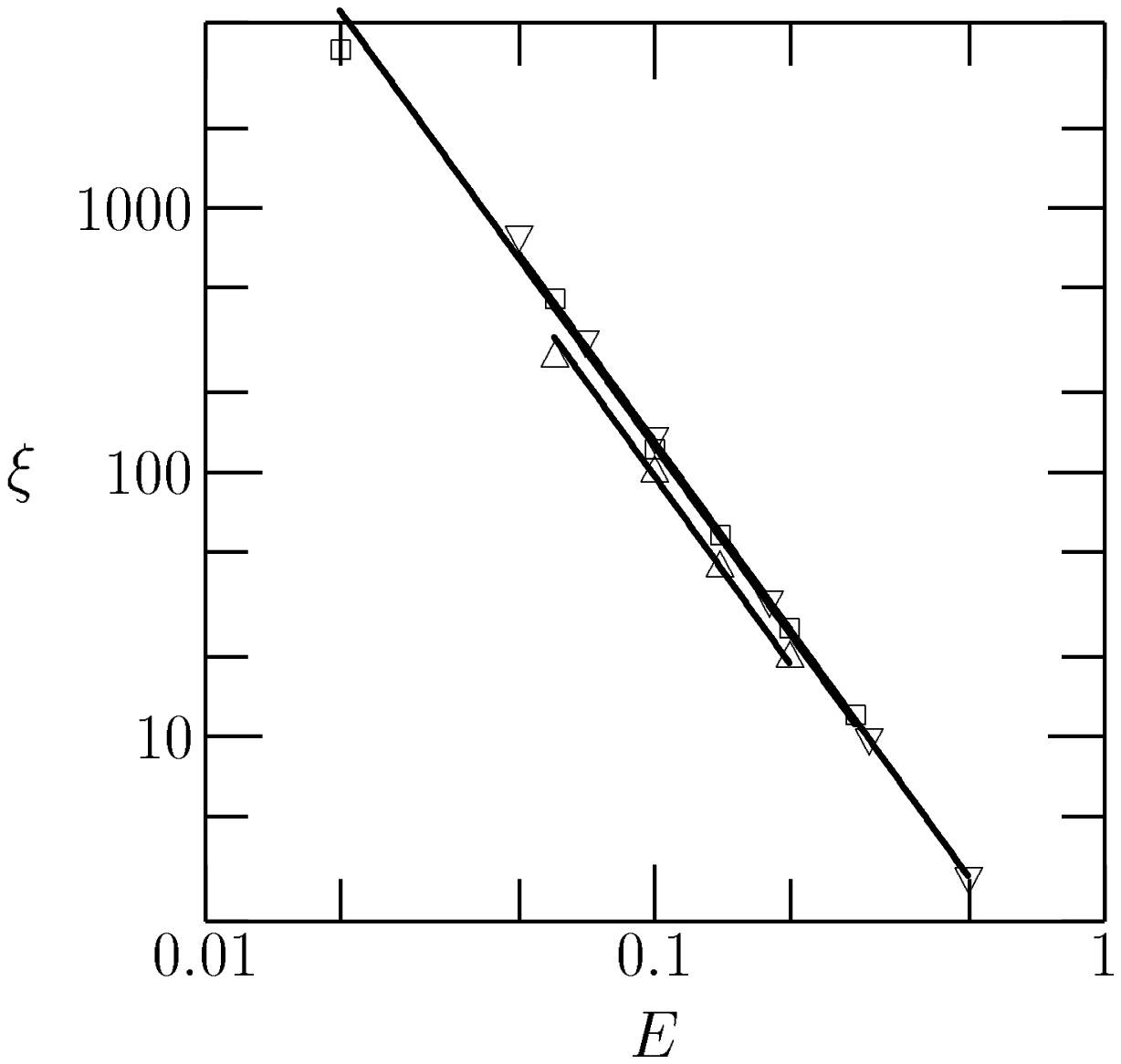}
\caption{Localization length $\xi$ (units $\ell_B$) as a function of 
  the energy $E$ (units $\Gamma$) for $n=0$, $\sigma=0$ ($\triangledown$);
  $n=0$, $\sigma =\ell_{B}$ ($\triangle$); $n=1$, $\sigma = \ell_{B}$ ($\Box$)
  (from \protect\cite{h92}).}
  \label{kvkfig:2}
\end{figure}

Very careful quantitative determinations of the critical exponent of such
Hamiltonians have been carried out \cite{hk90} by using the numerical scaling
method introduced earlier \cite{m80,mk81,mk83,km93}. The results are described
in detail in \cite{h95}. It was found that the value of the exponent is
universally $\nu=2.35\pm 0.03$ which is clearly different from the value
$2.73\pm 0.02$ for the 2D symplectic case \cite{aso02}. This is up to now considered as the most
reliable value which is remarkably close to an earlier non-rigorous suggestion
$\nu=7/3$ \cite{ms88} which was obtained on the basis of a semiclassical
percolation model augmented by a quantum tunneling argument (see below). In
the lowest and the second lowest band for white noise potential and finite
correlation length the same exponent was obtained within the numerical
uncertainty (Fig.  \ref{kvkfig:2}). Furthermore, the exponent also turned out
to be independent of long range correlations in the impurity potential (see
below). Thus, to the best of our knowledge, the QHT represents a quantum phase
transition for which it has been possible within the numerical uncertainties
explicitly to demonstrate universality over a wide range of parameters for the
first time.

\subsection{The Peierls model}
One of the standard models of localization is the Anderson Hamiltonian for a
disordered system
\begin{equation}
  \label{eq:05}
  H=\sum_{j}\epsilon_{j}|j\rangle\langle j|+\sum_{[jk]}V_{jk}
|j\rangle\langle k|\,.
\end{equation}
Here, $j$ denote the sites of a hypercubic lattice, $\sum_{j}|j\rangle\langle
j|=1$ are the complete set of corresponding site states, $\epsilon_{j}$ the
random site energies distributed randomly and independently in the interval
$[-W/2,W/2]$, and $V_{jk}$ the (real) hopping matrix elements between nearest
neighbors that are in the simplest case assumed to be a constant $V$.  This
model often has been used in numerical studies of the localization problem in
$d$ dimensions without magnetic field \cite{mk93}. In 2D, without disorder,
the energy spectrum covers the range $-4V<E<4V$. With disorder, the region of
eigenenergies is $-4V-W/2<E<4V+W/2$, and all states are localized.

With magnetic field, the hopping matrix elements acquire Peierls phase factors
\cite{p33,l51}
\begin{equation}
  \label{eq:06}
  V_{jk}=V\,\exp{\left\{-\frac{ie}{\hbar}\int_{j}^{k}{\rm d}{\bf r}\cdot
{\bf A}({\bf r})\right\}} := V e^{i\alpha_{jk}}\,.
\end{equation}
In the absence of disorder, the energy spectrum shows very rich and nontrivial
self-similarity features that depend on the commensurability between the
length scale imposed by the magnetic field and the lattice constant $a$. This
is the ''Hofstadter butterfly'' \cite{h76}. If the number of flux quanta per
unit cell, $n_{B}a^{2}:=\alpha$ is rational, $\alpha=p/q$, the spectrum
consists of $q$ absolutely continuous ´bands of eigenenergies. If $\alpha$ is
irrational, the spectrum is singularly continuous, it consists of a dense
distribution of isolated eigenvalues. It can be shown by using the effective
mass approximation that the spectrum becomes Landau like near the band edges.

Using the Peierls tight binding model, numerical studies showed delocalization
in the centers of the bands \cite{smk84} and quantization of the Hall
conductivity in the localized regime \cite{mk85}.

\subsection{The random network model}
The origin of the randomness in a quantum Hall sample is mainly the random
potential induced by the impurities in the bulk of the semiconductor sample
due to the doping. In AlGaAs heterostructures the inversion layer is far from
the doping layer. Therefore, the potential landscape seen by the electrons is
rather smooth. Spatial correlations can be expected to be important when
aiming at describing experimental results. Also from the theoretical viewpoint
it turns out to be very useful to consider spatially long-ranged randomness.
This is related to the fact that if the potential is smooth on the length
scale of the magnetic field the eigenvalue problem can be discussed in terms
of semiclassical approximation and the determination of the eigenstates
becomes equivalent to a percolation problem \cite{t76,i82,kl82,o82a,pj82,t83}.

If the correlation length of the random potential is large as compared to the
magnetic length, $\sigma\gg \ell_{B}$, it can be shown that the dynamics of
the electrons are governed by two totally different lengths scales. This limit
can always be achieved if the magnetic field is assumed to be sufficiently
large. In this limit, the eigenenergies are given by the equipotential lines,
\begin{equation}
  \label{eq:07}
  E=\left(n+\frac{1}{2}\right)\hbar \omega_{B}+V(X,Y)\,,
\end{equation}
with center coordinates $(X,Y)$. The corresponding wave functions represent a
slow drift of probability along the equipotential lines (length scale
$\sigma$) of electrons that perform rapid cyclotron motions around $(X,Y)$
(length scale $\ell_{B}$). The direction of the probability current is
perpendicular to the magnetic field and perpendicular to the local electric
field represented by the gradient of the potential along the equipotential
line. Reversing the direction of the magnetic field reverses the direction of
the current. Formally, the wave functions can be considered as waves
superposed of the Landau states associated with the equipotential lines.  The
extension of the wave functions perpendicular to the equipotential lines is of
the order of $\ell_{B}$. Formally, one can approximate \cite{t83}
\begin{equation}
  \label{eq:08}
\psi(u,v)= C(u)\chi(v)e^{i\phi(u,v)} 
\end{equation}
with $u$ and $v$ the coordinates parallel and perpendicular to the
equipotential line, respectively, $C^{-2}(u)$ the local electric field at
$(u,v=0)$ and $\phi(u,v)$ a gauge dependent phase. In this semiclassical
limit, the allowed energies are determined from the condition that $\phi(u,v)$
must change by an integer multiple of $2\pi$ when moving around a closed
equipotential contour.

\begin{figure}[htbp]
  \begin{center}
\includegraphics[width=0.6\linewidth]{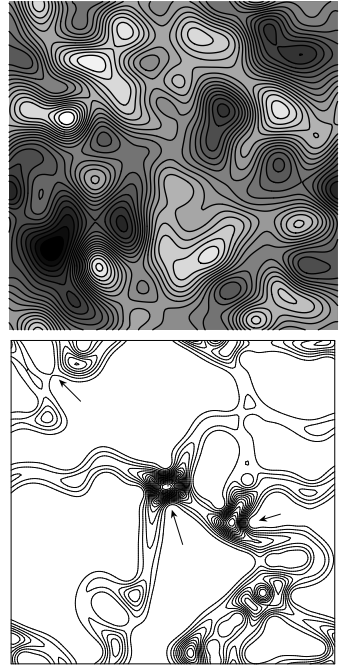}
\hspace{1mm}
\includegraphics[width=0.57\linewidth]{wavecont3.ps}
  \end{center} 
 \caption[]{Top: gray scale picture of a correlated random
    potential, $\sigma=2\ell_{B}$, with equipotential lines, white: high,
    black: low potential; bottom: modulus of a wave function corresponding to
    an energy near the center of the Landau band extending essentially along
    equipotential lines, occasionally inter-connected via tunneling near the
    saddle points of the potential (arrows).}
  \label{kvkfig:3}
\end{figure}

The average total spatial extend of a given wave function is determined by the
mean diameter of the randomly percolating equipotential line. When at some
point the distance between two equipotential lines gets closer than $\ell_{B}$,
quantum tunneling will lead to coupling between corresponding wave functions.
These regions correspond to saddle points of the potential.  They are
especially important near the center of the band (Fig.~\ref{kvkfig:3}) where
the critical behavior of the localization properties is expected to be
determined by the competition between quantum interference and tunneling.

The percolation picture allows immediately to draw two important conclusions.
First, at energies far away from the band center, corresponding to states in
the tails of the band, the equipotential lines are closed percolating
trajectories. Here, the wave functions are localized. Moving the energy
towards the center of the band from above or below will increase the mean
diameter of the percolating equipotential line. At some critical energy
$E_{\rm c}$ near the band center, the equipotential line will percolate
throughout the whole system (Fig.~\ref{kvkfig:3}). This is the percolation
threshold. For a symmetric potential distribution, this happens exactly at the
band center. Thus, there is at least one energy, where the ''classical''
localization length --- the mean diameter of the percolating cluster ---
becomes of the order of the size of the system. Second, according to
percolation theory, the mean diameter of a connected cluster near the
threshold diverges according to a power law, $\xi_{\rm p}\propto |E-E_{\rm
  c}|^{-\nu_{\rm p}}$ \cite{s79}, with the percolation critical exponent
$\nu_{\rm p}=4/3$ \cite{n79,be81}. It has been suggested \cite{ms88} that
tunneling across the saddle points supports the delocalization near the
percolation threshold. In particular, it was argued that the localization
exponent is increased exactly by 1, $\gamma=\nu_{\rm p}+1=7/3$.

\begin{figure}[htbp]
  \begin{center}
    \includegraphics[width=0.6\linewidth]{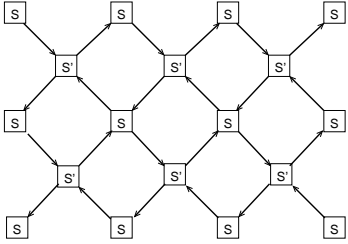}  
  \end{center}  
\caption{Bipartite network of saddle points S, S' connected by links 
  (after \protect\cite{cc88}). Directions of the probability currents in the
  links are indicated by arrows. Current directions change from clockwise to
  couter clockwise between neighboring loops in the diagonals.}
  \label{kvkfig:4}
\end{figure}

Besides estimating the critical exponent, the semiclassical limit provides the
background of a standard model for the physics of a system in the quantum
Hall regime near the quantum critical point \cite{cc88}.  Basically, one
replaces the irregular assembly of different saddle points of the
semiclassical percolation network (Fig.~\ref{kvkfig:3}) by a regular network
of saddle point scatterers connected by links which carry the scattering wave
functions associated with the above equipotential lines far away from a saddle
point (Fig.~\ref{kvkfig:4}).

Each saddle point has attached four links carrying amplitudes $\psi_{1}\ldots
\psi_{4}$, two of them, say $\psi_{1},\psi_{3}$, describing incoming waves and
the other two outgoing waves. As a result of the magnetic field incoming and
outgoing waves are spatially separated, in contrast to a normal transmission
problem. By solving the scattering problem for a saddle point
$V(x,y)=E_{0}-ax^{2}+by^{2}$ exactly \cite{f88}, the transmission probability
at energy $\epsilon:=(E-E_{0})/\ell_{B}^{2}(ab)^{1/2}$ was found,
$|t|^{2}=1/[1+\exp{(-\pi\epsilon})]$.

In terms of the reflection and transmission amplitudes, $r$ and $t$,
respectively, the scattering properties of the saddle point can be
parametrized as
\begin{equation}
  \label{eq:09}
\left(  
\begin{array}{c}
\psi_{2}\\
\psi_{4}
  \end{array}
\right)
={\bf S}
\left( 
\begin{array}{c}
\psi_{1}\\
\psi_{3}
  \end{array}
\right)
\end{equation}
with the scattering matrix
\begin{equation}
  \label{eq:10}
  {\bf S}=
\left(
  \begin{array}{cc}
e^{-i\phi_{2}}&0\\
0&e^{i\phi_{4}}
  \end{array}
\right)
\left(
  \begin{array}{cc}
-r&t\\
t&r
  \end{array}
\right)
\left(
  \begin{array}{cc}
e^{i\phi_{1}}&0\\
0&e^{-i\phi_{3}}
  \end{array}
\right)
\end{equation}
with $|r|^{2}+|t|^{2}=1$. The phases $\phi_{j}$ ($j=1\ldots 4$) correspond to
the waves in the links. At ${\bf S'}$ (Fig.~\ref{kvkfig:4}), the role of
incident and outgoing channels is interchanged.

Exactly at the energy of the saddle point, $r(E_{0})=t(E_{0})=1/\sqrt{2}$. If
$E\ll E_{0}$, $r\approx 1$, and the incoming waves $\psi_{1}$, $\psi_{2}$ are
completely reflected into outgoing waves $\psi_{2}$, $\psi_{4}$, respectively.
If $E\gg E_{0}$, $t\approx 1$. In this case, $\psi_{1}$, $\psi_{2}$ are
transmitted into $\psi_{4}$, $\psi_{2}$, respectively. These latter energy
regions correspond to wave functions localized completely within the loops of
the network --- equivalent to the valleys and the mountains of the potential
landscape --- while in the former case the wave functions can extend across
the whole network.

By numerically calculating the transmission through a network of identical
saddle points but with random phases in the links, the critical behaviour of
the localization length has been determined. The critical exponent in the
lowest Landau band was found to be consistent with the values obtained earlier
\cite{cc88}.

In order to better understand the critical point, the real-space
renormalization group approach has been adapted to describe the critical point
of the Chalker-Coddington network model \cite{gr97,ajs97,jmmw98,cr03}.  The
results obtained for the exponent are in good agreement with the earlier
findings. In summary, it appears now that there is more than convincing
evidence for the picture of the QHT as a universal quantum phase transition.
In the next section, we provide an overview of the experimental evidences that
this quantum phase transition is indeed physically realized in the QHE
experiments.

\subsection{The Dirac model}

The above network model of Chalker and Coddington is defined in terms of a
scattering problem {\em at a given energy}. This is of great practical
usefulness for studying the localization critical behaviour. By reconstructing
from the transfer matrix the underlying Hamiltonian one can arrive at a class
of models that allows to relate the QHT to other quantum critical phenomena.

\begin{figure}[htbp]
  \begin{center}
    \includegraphics[width=3.5cm]{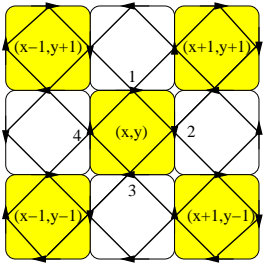}
  \end{center}
  \caption{The network model re-arranged as a tight binding 
    model on a 2D square lattice with four basis states per ''site'' (grey
    squares) at position $(x,y)$ ($x+y={\rm even}$). The links between the
    saddle points in Fig.~\protect\ref{kvkfig:4} (indicated here by
    $\diamondsuit$) are associated with the basis states (arrowheads). They
    are connected by tunneling (thick solid lines) across the saddle points.}
  \label{kvkfig:5}
\end{figure}

First, we establish a connection between the network model with a {\em tight
  binding model}. We re-arrange the network in the way shown in
Fig.~\ref{kvkfig:5}. The loops that are inter-connected by the saddle points
are arranged as a 2D lattice such that the centres of the loops are associated
with lattice points ${\bf R}_{xy}=x{\bf e}_{x}+y{\bf e}_{y}$ with integer
$x,y$ and $x+y={\rm even}$. The $x$- and $y$-directions are assumed
parallel to the directions of the links that connect $S$ and $S'$
(Fig.~\ref{kvkfig:4}). The links between adjacent saddle points within each of
the loops (denoted by arrowheads labelled with $\lambda=1,\ldots,4$ in
Fig.~\ref{kvkfig:5} \cite{hc96}) are associated with four ''site'' states.
These are then characterized by the lattice vector and the ''quantum number''
$\lambda$. They are assumed to form a complete set and are connected within a
given unit cell and between nearest neighboring cells via tunneling across the
saddle points.
 
The matrix elements of the effective Hamiltonian can be determined following
\cite{hc96,km95}. The vector of the amplitudes at ${\bf R}_{xy}$ after $M+1$
iterations is
\begin{equation}
  \label{eq:timeevolution}
  \psi_{{\bf R}\lambda}(M+1)=
\sum_{{\bf R}'\lambda'=1}^{4}U_{{\bf R}\lambda,{\bf R}'\lambda'}
\,\psi_{{\bf R}'\lambda'}(M)\,.
\end{equation}
The unitary operator ${\bf U}$ describes the evolution of the wave function
between $M$ and $M+1$. The eigenstates of the $4L$-dimensional matrix ${\bf
  U}$ ($L$ system size)
\begin{equation}
  \label{eq:unitary}
  {\bf U}\,|\,\psi_{\alpha}\rangle=e^{i\omega_{\alpha}(E)}\,
|\psi_{\alpha}\rangle\qquad(\alpha=1\ldots\ldots4L)
\end{equation}
with eigenvalues 1 ($\omega_{\alpha}(E)=0$) are the stationary states of the
network with an energy parametrized by the energy dependent reflection
parameter of the saddle points, $r\equiv r(E)=\sqrt{1-t^{2}(E)}$.

Interpretation of (\ref{eq:timeevolution}) as a ``time dependent''
Schr\"o\-din\-ger equation, $\psi(M+1)-\psi(M)=({\bf U}-{\bf 1})\psi(M)$,
suggests to relate ${\bf U}$ to a self-adjoint Hamiltonian ${\bf H}$,
\begin{equation}
  \label{eq:hamiltonian}
  {\bf H}:=\frac{1}{2i}\left({\bf U}^{\dagger}-{\bf U}\right)\,,
\end{equation}
with quasi-energy eigenvalues $\epsilon_{\alpha}=\hbar \omega_{\alpha}(E)$.

Since each lattice point is connected via four saddle points to its nearest
neighbors, ${\bf U}$ has a $4\times 4$ block structure. Due to the bipartite
structure of the links each of the $4\times 4$ blocks has a $2\times 2$
block structure. By suitably arranging amplitudes one gets
\begin{equation}
  \label{eq:twoblocks}
  {\bf U}=
\left(
  \begin{array}{cc}
0&{\bf M}\\{\bf N}&0
  \end{array}
\right)
\end{equation}
which gives for the effective Hamiltonian
\begin{equation}
  \label{eq:twoblocks(h)}
   {\bf H}=
\frac{1}{2i}\left(
  \begin{array}{cc}
0&{\bf N}^{\dagger}-{\bf M}\\{\bf M}^{\dagger}-{\bf N}&0
  \end{array}
\right)
\,.
\end{equation}
The matrices
\begin{equation}
  \label{eq:m}
 {\bf M}=
\left(
  \begin{array}{cc}
te^{i\phi_{1}}\tau_{-}^{x}\tau_{+}^{y}&re^{i\phi_{1}}\\
re^{i\phi_{3}}&-te^{i\phi_{3}}\tau_{+}^{x}\tau_{-}^{y}
  \end{array}
\right)\,,
\end{equation}
\begin{equation}
  \label{eq:n}
 {\bf N}=
\left(
  \begin{array}{cc}
re^{i\phi_{2}}&te^{i\phi_{2}}\tau_{+}^{x}\tau_{+}^{y}\\
te^{i\phi_{4}}\tau_{-}^{x}\tau_{-}^{y}&-re^{i\phi_{4}}
  \end{array}
\right),
\end{equation}
are obtained using the scattering matrix ${\bf S}$. They contain translation
operators $\tau_{\pm}$ connecting neighboring cells
\begin{equation}
  \label{eq:translations}
  \tau_{\pm}^{x}\psi_{{\bf R}\lambda}=\psi_{{\bf R}\pm {\bf e}_{x}\lambda}
\qquad
\tau_{\pm}^{y}\psi_{{\bf R}\lambda}=\psi_{{\bf R}\pm {\bf e}_{y}\lambda}\,.
\end{equation}
Inserting this into the Hamiltonian one notes that
within a given cell, say at ${\bf R}$, the Hamiltonian matrix
elements are proportional to the reflection amplitudes
\begin{equation}
  \label{eq:intracell}
  h_{{\bf R}j,{\bf R}j-1}=\frac{ir}{2}\,e^{i\phi_{j}},
\end{equation}
with $j=1,2,3,4$ cyclic (such that $j-1:=4$ for $j=1$). The eight matrix
elements that couple nearest neighbor cells contain the transmission amplitude
$t$,
\begin{eqnarray}
  \label{eq:intercell}
  h_{{\bf R}2,{\bf R}_{++}3}=\frac{it}{2}\,e^{i\phi_{2}},\quad&&
  h_{{\bf R}3,{\bf R}_{+-}4}=\frac{-it}{2}\,e^{i\phi_{3}},\\
  h_{{\bf R}4,{\bf R}_{--}1}=\frac{it}{2}\,e^{i\phi_{4}},\quad&&
  h_{{\bf R}1,{\bf R}_{-+}2}=\frac{it}{2}\,e^{i\phi_{1}}\,,
\end{eqnarray}
with ${\bf R}_{\pm\pm}:={\bf R}+(\pm 1,\pm 1)$ (Fig.~\ref{kvkfig:5}). The
remaining matrix elements are the conjugates of these.

When the phases $\phi_{j}$ are independent of the lattice point ${\bf R}$, and
all of the saddle points are identical, the system is periodic. The
Hamiltonian can be diagonalized exactly with a Bloch Ansatz
\begin{equation}
  \label{eq:bloch}
  \psi_{\lambda}({\bf R})=e^{i{\bf q}\cdot{\bf R}}u_{\lambda}(\bf q)\,.
\end{equation}
The resulting band structure $\epsilon({\bf q})$ can be straighforwardly
obtained in a closed form. It consists of two bands separated by a gap
$\Delta$ at ${\bf q}=0$. This can be seen most easily by considering the
eigenvalue problem for ${\bf H}^{2}$ instead of ${\bf H}$. Exactly at the
energy of the (identical) saddle points, $r=t=1/\sqrt{2}$ and the gap
vanishes, $\Delta =0$.  For energies close to the saddle point, one expands
$r:= 1/\sqrt{2}+\Delta/4$ ($\Delta\ll 1$). One finds in this case $\Delta=
2\sqrt{(1-2rt)}$ when assuming $\sum_{j=1}^{4}\phi_{j} =\phi=\pi$ wich is
equivalent to a flux per lattice site of $\Phi=\hbar \phi/e=h/2e$.

The formation of the gap can be understood in better detail by expanding the
band structure near ${\bf q}\approx 0$. One finds for small $\Delta$
\begin{equation}
  \label{eq:effmassdirac}
  \epsilon^{2}({\bf q})=\Delta^{2}+({\bf q}-e{\bf A})^{2}
\end{equation}
with a ''vector potential'' ${\bf
  A}=(1/2)\left(\phi_{1}-\phi_{3},\phi_{4}-\phi_{2}\right)$.
In this ''effective mass approximation'', one can show that the Hamiltonian
  has the Dirac form  
\begin{equation}
  \label{eq:dirac}
  {\bf H}=(p_{x}-eA_{x}){\bf \sigma}_{x}+(p_{y}-eA_{y}){\bf \sigma}_{y}+
m{\bf \sigma}_{z}+\phi{\bf 1}\,,
\end{equation}
with the components of the momentum operator $p_{j}=-\hbar i\partial_{j}$
($j=x,y$),  the mass $m\equiv \Delta$, and the Pauli
matrices ${\bf \sigma}_{1}$, ${\bf \sigma}_{2}$, ${\bf \sigma}_{3}$.

In this Hamiltonian, randomness can be introduced in different ways. Via
randomness in the individual phases one can make the components of the vector
potential random. Randomizing the total Aharonov-Bohm phases in the loops
produces randomness in the ''scalar potential'' $\phi$. Finally, assuming the
tunneling parameters of the saddle points to be random gives fluctuations in
the mass parameter $m$.

The correspondence between the quantum Hall problem, certain tight binding
models, and the 2D Dirac model has been noted by several authors
\cite{ff85,f86,lfsg94,l94,z95}. Fisher and Fradkin \cite{ff85} have reached
the Dirac model starting from a 2D tight binding model with diagonal on-site
disorder in a perpendicular magnetic field with half a flux quantum per unit
cell. They constructed a field theory for the diffusive modes which was shown
to be in the same universality class as the orthogonal $O(2n,2n)/O(2n)\times
O(2n)(n\to 0)$ non-linear $\sigma$-model. This implies that all states are
localized as in the absence of a magnetic field and suggests that if
delocalization occurs with magnetic field, it must be a direct consequence of
breaking time-reversal symmetry instead of some other property of the field.
Generalizations to the several-channel scattering problem have also been
discussed \cite{f86}.

\begin{figure}[htbp]
 \begin{center}
  \includegraphics[width=4cm]{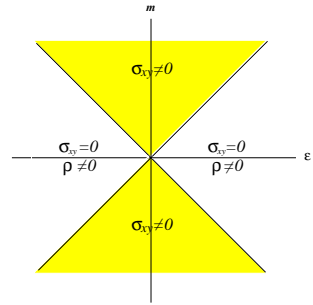}    
    \caption{The phase diagram of the ordered Dirac model 
      \protect\cite{lfsg94} showing regions of non-zero density of states
      $\rho$ ($|\epsilon|>|m|$) and non-zero Hall conductivity
      $\sigma_{xy}=\pm\,e^{2}/2h$ ($|\epsilon|<|m|$).}
    \label{fig:dirac}
  \end{center}
\end{figure}

Ludwig and collaborators \cite{lfsg94} have used a tight binding model on a
square lattice with nearest and next-nearest neighbor coupling, half a flux
quantum per unit cell and a staggered potential energy $\mu (-1)^{x+y}$ as a
starting point. At low energy, this model was shown to be equivalent to a
Dirac model with two Dirac fields. Without disorder the model exhibits an IQHE
phase transition as a function of a control parameter which is essentially
the mass $m$ of the lighter Dirac field. The transition belongs to the
2D-Ising universality class. The associated exponents and the critical
transport properties were determined. The density of states is
\begin{equation}
  \label{eq:diracdos}
  \rho(\epsilon)=\frac{|\epsilon|}{\pi}\Theta(\epsilon^{2}-m^{2})\,.
\end{equation}
It can readily be obtained from (\ref{eq:dirac}) with ${\bf A}=0$.
Applying linear response theory to the Dirac system one can determine the Hall
conductivity by calculating the ratio of, say, the current density in the
$x$-direction, $j_{x}$ and the electric field, $E_{y}$, in the $y$-direction
\begin{equation}
  \label{eq:sigmaHall}
\sigma_{xy}\equiv \frac{j_{x}}{E_{y}}=\frac{e^{2}}{h}\,\frac{m}{2\pi^{2}}\,
\int\frac{{\rm d}^{2}q {\rm d}\omega}
{[(i\omega-\epsilon)^{2}-q^{2}-m^{2}]^{2}}\,.
\end{equation}
It is found that at zero energy, $\epsilon =0$, where the above density of
states vanishes, the Hall conductivity jumps by $e^{2}/h$ at $m=0$
(Fig.~\ref{fig:dirac})
\begin{equation}
  \label{eq:diracHall}
  \sigma_{xy}(m)=-\frac{{\rm sgn}(m)}{2}\,\frac{e^{2}}{h}\,.
\end{equation}
The heavier Dirac field contributes towards the Hall conductivity with
$e^{2}/2h$ such that the total Hall conductivity jumps from $0$ to $e^{2}/h$.
Simultaneously, the magneto-conductivity $\sigma_{xx}$ is non-zero,
\begin{equation}
  \label{eq:sigmaxx}
  \sigma_{xx}=\sigma_{0}\frac{e^{2}}{h}\delta_{m,0}\,,
\end{equation}
with the Kronecker-symbol $\delta_{m,0}$ equal to 1 for $m=0$ and 0 for $m\neq
0$, and the constant $\sigma_{0}$ of the order $\pi/8$. The critical point
shows time-reversal, particle-hole and parity invariance.
Thus, the clean 2D Dirac model exhibits a QHT at
$\epsilon=m=0$.  By introducing randomness in the Dirac mass, $m=m(x,y)$, the
wave functions are confined to the contours $m(x,y)=0$. Then, the QHT can be interpreted in terms of quantum percolation of these states.
In the absence of randomness of the phases (i.e. the vector potential), the
corresponding critical exponent of the correlation length is that of the
classical percolation model.

Introducing disorder in the various parameters of the model leads to breaking
of the symmetries, as indicated in Table~\ref{tab:diracsymmetries}.
\begin{table}[htbp]
\vspace{1mm}
\caption[]{Symmetry breaking by disorder in the various kinds of parameters of
  the Dirac model (${\bf p}$ momentum operator; $m$ Dirac mass parameter;
  $\sigma$ vector of the Pauli matrices $\sigma_{x}$, $\sigma_{y}$,
  $\sigma_{z}$; $\phi$  scalar potential; ${\bf A}$ vector potential).}
\label{tab:diracsymmetries}
\begin{center}
\vspace{1mm}
\begin{tabular}{lcccc}
\hline\hline
symmetry&${\bf \sigma}\cdot{\bf p}$&$m\sigma_{z}$&$\phi$
&${\bf \sigma}\cdot{\bf A}$\\\hline
parity&yes&no&no&no\\
time reversal&yes&no&yes&no\\
particle-hole&yes&no&no&yes\\
\hline
\end{tabular}
\end{center}
\end{table}
Taking into account only randomness in the scalar potential, it was argued
that the transition can be described by a symplectic non-linear sigma
model.

The case of only a random vector potential can be treated to a large extend
analytically \cite{ac79,ntw94,b95,ckt96,kmt96} and it has many remarkable
properties. The model exhibits a fixed line in this case. The zero-energy wave
function was exactly determined. It was found to be extended and showed
multifractal scaling. The density of states was found to show a power law
dependence on the energy near $\epsilon\approx 0$ with the exponent varying
continuously upon moving along the fixed line while the diagonal conductivity
was found to be constant. With all three different types of disorders
included, the Dirac model is equivalent to the Chalker-Coddington network
model with randomness in the saddle points included.

\section{One-parameter scaling}

Table \ref{critexp} contains a representative selection of exponents
numerically obtained so far from numerical investigations done on different
models. The question arises how the QHT, i.e its critical properties can be
investigated experimentally.
\begin{table}[htbp]
\caption{Selection of critical exponents in the lowest Landau
  band. 
Abbreviations: 
CCN (Chalker-Coddington network model), 
CNS (Chern number scaling), 
DLS (double layer system with white noise randomness), 
FRI (finite range impurities), 
PTB (Peierls tight binding Hamiltonian), 
RGF (recursive Green function method), 
RLM (random Landau matrix model), 
RSN (random saddle point network model),
SSC (superspin chain) 
RSR (real space renormalization),
SCD (self-consistent diagrammatic perturbation theory), 
SOS (spin orbit scattering), 
SRI (short range impurities), 
TNS (Thouless number scaling), 
TMS (transfer matrix scaling), 
FSS (finite size scaling)
}
\label{critexp}
\vspace{1mm}
\begin{tabular}{p{2.5cm}p{1.8cm}p{1.8cm}l}\hline\hline
  exponent $\nu$  & model & method&reference\\\hline
  $\infty$        &SRI&SCD&\protect\cite{o82}\\
  $\approx 2$     &PTB&TMS&\protect\cite{smk84}\\
  $\approx 2.0$   &SRI&RGF&\protect\cite{aa85}\\
  $2.35\pm 0.03$  &RLM&RGF&\protect\cite{hk90,h95,aso02}\\
  $2.3\pm 0.08$   &RLM&RGF&\protect\cite{m90}\\
  $2.4\pm0.2$     &RLM&RGF&\protect\cite{h92}\\
  $2.4\pm0.1$     &FRI&CNS&\protect\cite{hb92}\\
  $\approx 2.3$   &SOS&TNS&\protect\cite{hamg95}\\
  $\approx 2$     &DLS&TNS&\protect\cite{sm96}\\
  $2.5\pm 0.5$    &CCN&TMS&\protect\cite{cc88}\\
  $2.43\pm 0.18$  &RSN&TMS&\protect\cite{lwk93}\\
  $2.5\pm 0.5$    &CCN&RSR&\protect\cite{gr97}\\
  $2.39\pm0.01$   &CCN&RSR&\protect\cite{cr03}\\
  $2.38\pm 0.4$   &SSC&FSS&\protect\cite{kondev}\\
\hline\hline
\end{tabular}
\end{table}

\subsection{The scaling picture}
As the QHT is a quantum phenomenon in an {\em infinite} system at zero
temperature, it is obvious that for determining the critical behavior some
extrapolation method is required. This has been developed
\cite{mk81,mk83,km93} by starting from the finite size scaling method
previously established for conventional phase transitions \cite{b83}.

One defines a quantity $X$ which depends on a set of parameters ${x_i}$
characterizing the system, such as the Fermi energy, the variance of the
disorder, the magnetic field, and the system size $L$,
\begin{equation}
X=f(\{x_i\},L) .
\end{equation}
We assume the existence of a scaling law 
\begin{equation}
X=F(\chi L^{1/\nu},\phi_1 L^{y_1},\phi_2 L^{y_2},\cdots)\, ,
\label{eq:scaling}
\end{equation}
with $\chi$ being the relevant scaling variable and $\phi_i$ denote the
irrelevant ones. The latter nevertheless can cause corrections to scaling as
long as the system size is finite. In a numerical calculation they must not be
ignored. These variables characterize distances from the critical point,
$\nu>0$ is the critical exponent and $y_i<0$ are the irrelevant exponents.
For large enough system size, only the relevant scaling variable survives, and
(\ref{eq:scaling}) becomes
\begin{equation}
X=F_1\left(\frac{L}{\xi}\right) ,
\label{eq:fss}
\end{equation}
with $\xi\sim \chi^{-\nu}$. The quantity $\chi$ as a function of some control
parameter, say $x$, can be expanded near the critical point $x_{\rm c}$
\begin{equation}
\chi=a_1 (x-x_{\rm c})+a_2 (x-x_{\rm c})^2+\cdots .
\label{eq:chi_expansion}
\end{equation}
For $L\to\infty$, the scaling variable $X$ should show a singularity at the
localization-delocalization critical point. Most conveniently, the
localization lengths $\lambda(L; x)<\infty$ of a quasi-1D system can be used,
with $L$ the cross sectional system diameter \cite{mk81}. Consider now
\begin{equation}
\Lambda(L; x):=\frac{\lambda(L; x)}{L} \,.
\end{equation}
For parameters such that $\Lambda(L\to\infty; x)\to 0$ the system is localized
with the localization length $\xi(x):=\lambda(L\to\infty; x)<\infty$. If
$\Lambda(L\to\infty; x)\to \infty$ the system is delocalized. Here,
$\lambda^{-1}(L\to\infty; x)$ corresponds to the dc-conductivity
\cite{mk81,mk83}. The critical point is defined by $\Lambda(L\to\infty;
x):=\Lambda_{{\rm c}}={\rm const}$. Most efficiently, $\lambda(L;x)$ can be
calculated using the transfer matrix method. The quasi-1D localization length
is then given by the inverse of the smallest of the Lyapunov exponents of the
transfer matrix \cite{km93}.

Instead of the localization length one can also consider as scaling variables
other physical quantities such as the conductance, the level statistics and
the statistics of the wave functions. However, up to now the most precise
determinations of the critical behavior have been achieved using the
localization lengths (Tab.~\ref{critexp}).

\subsection{Experimental scaling}
Guided by the above one-parameter scaling method, one can infer the behavior
of the localization length from the scaling behavior of the experimentally
determined conductivity components with temperature $T$, frequency $\omega$ or
the geometrical size of the sample. In energy regions where the localization
length exceeds some cutoff length $L_{\rm c}$, the wave functions appear
effectively as extended, and the longitudinal conductivity $\sigma_{xx}\neq
0$. In the localized regions, $\xi(E)< L$, $\sigma_{xx}\to 0$ for $T\to 0$.
Correspondingly, in the latter regions $\sigma_{xy}=const$ while in the former
$\sigma_{xy}$ shows steps between successive Hall plateaus. The widths of the
conductivity peaks and/or the width of the steps in the Hall conductivity
reflect the widths of the energy regions of the effectively extended wave
functions.

The cutoff length can have very different physical origins. For system size
$L\to \infty$, at finite temperature, phase breaking processes, as for
instance induced by electron-phonon  
or electron-electron scattering, lead to a
temperature dependent phase breaking length $L_{\phi}(T)\propto T^{-p/2}$.
Similiarly, for finite frequency $\omega\neq 0$, the phase breaking length is
$L_{\omega}\propto \omega^{-z}$. On the other hand, for $T\to 0$ and
$\omega\to 0$, it is eventually the geometrical size of the system which plays
the role of the cutoff. In summary,
\begin{equation}
  \label{eq:4}
  L_{\rm c}={\rm min}(L_{\phi},L_{\omega},L)\,.
\end{equation}
For instance, the temperature dependence of the peaks in the dc-conductivity
component $\sigma_{xx}$ and the width of the corresponding steps in
$\sigma_{xy}$ can then be estimated starting from the idea that the peak width
is related to the width if the energy interval $\Delta(T):=2E_{\rm c}(T)$ of
the effectively extended states determined by the condition
$L_{\phi}(T)=\xi(E_{\rm c})\propto |E_{\rm c}|^{-\nu}$. This yields
$\Delta(T)\propto T^{p/2\nu}:=T^{\kappa}$ \cite{wetal88}. Such a behavior has
indeed been found in many experiments with a non-universal exponent
$\kappa=0.42$ \cite{wetal88,ketal91,eetal93}. An extensive description of the
data collected until 1995, and their interpetation, is given in \cite{h95}.
Recent measurements including the frequency scaling \cite{ketal0y,f98,hetal02}
confirm the scaling picture. 

Thus, to the best of our knowledge one can state that the QHE is indeed a
manifestation at finite $T$ of a universal quantum phase transition at $T=0$.

\section{The quest for the critical theory}

The numerical and experimental evidence for the universality of the quantum
Hall transition raises expectations that a full characterization of this
quantum critical point will result in the analytical derivation of its
critical exponents and the scaling functions. Soon after the discovery of the
QHE a field theory has been derived from the microscopic
Hamiltonian of the random Landau model (\ref{eq:01}) \cite{pruiskentop},
and (\ref{eq:02}) \cite{ef,weiden}. It was shown to have two coupling
parameters $\sigma^0_{xx}$ and $\sigma^0_{xy}$, the longitudinal and Hall
conductance as defined on small length scales of the order of the elastic mean
free path $l$. This field theory is based on the theory of localization of
electrons in weakly disordered systems. 

In order to describe localisation, one needs to go beyond perturbation theory
in the disordered potential, and has to take full advantage of the symmetries
arising in the calculation of correlation functions of disordered systems.
This complication can be traced back to the fact that the disorder averaged
electron wave function amplitude $\langle\psi ({\bf x}, t)\rangle$ decays on
length scales on the order of $l$, and contains no information on
localisation. Rather, in order to capture quantum localisation, one needs to
consider higher moments of the wave function amplitudes, such as the impurity
averaged evolution of the electron density $n({\bf x},t) = \langle |\psi (
{\bf x}, t)|^2\rangle$. Thus, in a more formal language, a nonperturbative
averaging of products of retarded and advanced propagators, $ \langle G^R ( E)
G^A (E')\rangle$ is needed to obtain information on quantum localisation.  

In a useful anology to the study of spin systems, the field theoretical
approach contracts the information on localisation into a theory of Goldstone
modes $Q$, arising from the global symmetry of rotations between the
functional integral representation of the retarded propagator $G^R$ (''spin
up'') and the advanced propagator $G^A$ (''spin down''). The field theory can
either be formulated by means of the replica trick, where the $N$ replicas can
be represented either by $N$ fermionic or $N$ bosonic fields, yielding a
bounded or unbounded symmetric space, respectively, on which the modes $Q$ are
defined \cite{pruiskentop}.  

Because of the necessity and the difficulty to perform the limit $N
\rightarrow 0$ at the end of the calculation a more rigorous supersymmetric
field theory has been formulated. This technique represents the product of
Green functions $G^R(E) G^A(E')$ by functional integrals over two fermionic
and bosonic field components, composing a supersymmetric field vector $\psi$.
The supersymmetric representation enables one to perform the averaging over
the disorder potential as a simple Gauussian integral \cite{ef,weiden}. 

As a result of the averaging one obtains an interacting theory of the fields
$\psi$ containing an interaction term $\propto \psi^4$, where the {\it
  interaction strength} is proportional to the variance of the disorder
potential $W^2$ (2). This term can now be decoupled by introducing another
Gaussian integral over $Q-$matrices. Clearly, the field $Q$ should not be a
scalar, otherwise we would simply reintroduce the Gaussian integral over the
random potential $V$.  Rather, in order to be able to describe the physics of
localization, the field $Q$ should capture the full symmetry of the functional
integral representation of the correlation function. Therefore, the Gaussian
integral is chosen to be over a 4$\times$4 matrix $Q$ which itself is an
element of the symmetric space defined by the matrices $A$ that leave the
functional integral invariant under the transformation $\psi \rightarrow A
\psi$. In the supersymmetric formulation, this matrix consists of two blocks
of 2$\times$2 matrices whose parameter space consists of a compact (bounded)
and a noncompact (unbounded) sector. The off-diagonal blocks, so to say the
rotations between the compact and the noncompact sector, are then found to be
parametrised by Grassmann (fermionic) variables.

Correlation functions such as the electron density and the conductivity can
then be obtained from a partition function of these fields $Q$.  The
spatial variations of $Q$ are governed by the action
\begin{eqnarray} \label{exactfree}
S[Q] = && 
 \frac{\pi\hbar}{4\Delta \tau}
  \int \frac{{\rm d} {\bf x}}{ L^2} {\rm Tr}\, Q ^{2}({\bf x}) 
\nonumber \\ &&
 \qquad\qquad + \frac{1}{2} \int  {\rm d} {\bf x}\,
 \langle{\bf x} |{\rm Tr}\, \ln
 G ( \hat{x}, \hat{p} )| {\bf x} \rangle
\end{eqnarray} 
where
\begin{eqnarray}
G^{-1}(\hat{x},\hat{p} ) &=&\left[
\frac{(\omega + i \delta) \Lambda_3}{2} - 
\frac{(\hat{p}-q A )^2}{2 m^{*}} - V_0(\hat{x})\right.\nonumber\\
&&\nonumber\\
&&\qquad\qquad\qquad\qquad\qquad + \left.\frac{i\hbar}{2 \tau}
Q (\hat{x}) \right]
\end{eqnarray}
where $1/\tau$ is the elastic scattering rate, and $\Delta$ the mean level
spacing related to the variance of the disorder potential (2) according to $
W^2 = \Delta \hbar/2 \pi \tau$.  The 4$\times$4 matrix $\Lambda_3$ is the
diagonal Pauli matrix in the sub-basis of the retarded and advanced
propagators, $\delta > 0$ and $\omega = E-E'$ break the symmetry between the
retarded and advanced sector.  

It turns out that the physics of diffusion and localisation, which arises on
length scales much larger than the elastic mean free path $l$, is governed by
the action of the long wavelength modes of $Q$. Thus, one can simplify and
proceed with the analysis by expanding around a homogeneous solution of the
saddle point equation, $\delta S = 0$.  For $\omega =0$, this is
\begin{equation} \label{saddle}
 Q =  \frac{i}{\pi \nu} \langle {\bf x}|
\left[E - H_0 - V_0({\bf x}) + \frac{i}{2 \tau} \,Q \right]^{-1}
| {\bf x} \rangle. 
\end{equation}  
This saddle point equation is solved by $ Q_0 = \Lambda_3 P $, which
corresponds to the self consistent Born approximation for the self energy of
the impurity averaged Green function. At $\omega =0$, rotations $ U$ which
leave the action invariant yield the complete manifold of saddle point
solutions as $ Q = \bar{U} \Lambda_3 P U$, where $ U \bar{U} = 1$.

The modes which leave $\Lambda_3$ invariant can be factorized out, leaving the
saddle point solutions in this supersymmetric theory to be elements of the
semi-simple supersymmetric space $Gl(2 \mid 2)/[Gl(1 \mid 1) \times Gl (1
\mid 1)] $ \cite{weiden}.

In addition to these gapless modes there are massive longitudinal modes with
$Q^2 \neq 1$ which only change the short distance physics, and not the
physics of localisation. They can be integrated out \cite{pruiskentop,ef}.
Thus, the partition function reduces to a functional integral over the
transverse modes $U$.

The action at finite frequency $\omega$ and slow spatial fluctuations of $Q$
around the saddle point solution can be found by an expansion of the action
$S$. Inserting $ Q = \bar{U} \Lambda_3 P U$ into (\ref{exactfree}) and
performing the cyclic permutation of $U$ under the trace ${\rm Tr}$
\cite{pruiskentop} allows a simple expansion to first order in the energy
difference $\omega$ and to second order in the commutator $U [ H_0, \bar{U}
]$. The first order term in $U [ H_0, \bar{U} ]$ is proportional to the local
current. It is found to be finite only at the edge of the wire in a strong
magnetic field, due to the chiral edge currents. It can be rewritten as
\begin{equation}
S_{xy II} = - \frac{1}{8} 
\int {\rm d}x {\rm d}y  \,
\frac{\sigma^{0 II}_{xy}({\bf x})}{e^2/h}\,{\rm STr} (Q 
\partial_x Q \partial_y Q)  
\end{equation} 
where the prefactor is the nondissipative term in the Hall conductivity in
self consistent Born approximation \cite{ef}
\begin{equation}
\sigma^{0 II}_{xy} ({\bf x}) = -\frac{1}{\pi} 
\frac{ \hbar e^2}{m^2} \langle{\bf x}| ( x \pi_y - y \pi_x ) 
 {\rm Im} G^R_{E} |  {\bf r} \rangle, 
\end{equation}
where ${\bf \pi } =(\hbar/i) {\bf \nabla} - q {\bf A }$.  This field theory
has now the advantage that one can treat the physics on different length
scales separately: the physics of diffusion and localization is governed by
the action of spatial variations of $U$ on length scales larger than the mean
free path $l$.  That is why this field theory is often called diffusive
nonlinear sigma model.

 The physics on smaller length scales  is included in the
coupling parameters of the theory, which is identified in the above derivation
as correlation functions of Green functions in self consistent Born
approximation, being related to the conductivity by the Kubo-Greenwood
formula,
\begin{eqnarray}\label{kubo}
\sigma^0_{\alpha \beta } ( \omega,  {\bf x} )& =& \frac{\hbar}{\pi  L^2}
\langle {\bf x} 
| \pi_{\alpha}  G^R_{0 E} {\bf  \pi_{\beta} }   
G^A_{0 E+\omega} | {\bf x} \rangle. 
\end{eqnarray} 
The remaining averaged correlators, involve products $ G^R_{0 E} G^R_{0
  E+\omega}$ and $ G^A_{0 E} G^A_{0 E+\omega}$ and are therfore by a factor
$\hbar/\tau E$ smaller than the conductivity, and can be disregarded for small
disorder. Using the Kubo formula (\ref{kubo}), the action  of $Q$
simplifies to
\begin{eqnarray} \label{free2db}
S &=& \frac{h }{16 e^2}  
\int {\rm d} {\bf x} \sum_{i=x,y}  \sigma^0_{ii} (\omega=0,{\bf x})
 \,{\rm Tr} \left[ ( { \nabla_i} Q({\bf x} ))^2 \right] 
\nonumber \\ 
&&-  \frac{h}{8 e^2}   \int {\rm d} {\bf x}  
\sigma^0_{xy} (\omega=0,{\bf x}) 
\,{\rm Tr} \left[ Q \partial_x Q \partial_y Q \right]
\end{eqnarray}
where $ \sigma^0_{xy} (\omega=0,{\bf x}) = \sigma^I_{xy} (\omega=0,{\bf x}) +
\sigma^{II}_{xy} (\omega=0,{\bf x})$ and $ \sigma^I_{xy} (\omega=0)$ is the
dissipative part of the Hall conductivity in self consistent Born
approximation (\ref{kubo}).

The first term in this action yields localization in 2D electron systems,
signaled by the presence of a gap in the field theory. The second term could
not be obtained by any order in perturbation theory. It is of topological
nature. In 2D and for a homogenous Hall conductance it can be shown that this
term can take only discrete purely imaginary values,
\begin{eqnarray} 
\label{free2dbtop}
S_{\rm Top} &=&  2 \pi i   \frac{h}{ e^2}    
\sigma^0_{xy} n,
\end{eqnarray}
where the integers $ n $ count how often the field $Q({\bf x})$ is winding
around its symmetric space as it varies spatially in 2D.
  
\begin{figure}[htbp]
   \includegraphics[width=7.5cm]{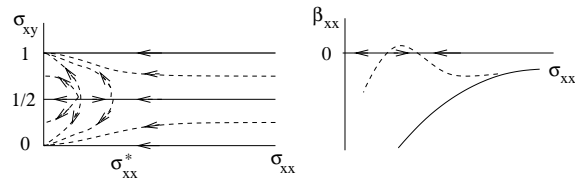}    
\caption[]{ The conjectured  two parameter flow diagram of 
  the integer quantum Hall effect (left), and the corresponding
  $\beta$--function, $\beta_{xx} = {\rm d} \ln \sigma_{xx}
/{\rm d} \ln L$ (right) at
  $\sigma_{xy} = 1/2$ (dashed line) and at $\sigma_{xy} = 0$ (full line).}
  \label{fig:flow}
\end{figure}

This theory was studied by means of a multi-instanton expansion, where one
sums over all solutions with different topological numbers $n$, of the saddle
point equation $\delta S = 0$, and integrates out fluctuations around these
instanton solutions on length scales exceeding $l$. Thereby one finds that
the conductance parameter $\sigma^0_{xx}$ becomes renormalized to smaller
values, but that this renormalization flow towards localization is slowed down
at half-integer values of the Hall conductance, $\sigma^0_{xy} =(n+1/2)
e^2/h$. This derivation is valid at large conductance parameters
$\sigma^0_{xx}$. On this basis the two-parameter scaling of the quantum Hall
transition has been suggested (Fig.~\ref{fig:flow}) \cite{pruiskentop,khm}.

Subsequently, it has been shown rigourously that this field theory is indeed
critical at half integer Hall conductance parameters $\sigma^0_{xy}$
\cite{pruiskentop,affleck,tsvelik}, and that it has a spectral gap to
fluctuations at other values of $\sigma^0_{xy}$. This indicates the
localization of the electron eigenstates of the random Landau model in the
tails of the Landau bands \cite{marikhin}.  Since the longitudinal conductance
at the critical point $\sigma^*$ is known to be smaller than $1$, the critical
point is located in the strong coupling limit of the field theory.  Thus, it
is outside of the validity of available analytical methods which can be used
to extract quantitative information on the critical exponents.

Recently, there has been nevertheless major progress, since it has been proven
that the Hamiltonian of a chain of antiferromagnetically interacting
superspins is equivalent to the nonlinear sigma model for short ranged
disorder at $\sigma_{xy} =1/2$ \cite{zirnchain}, as well as to the
Chalker-Coddington model \cite{lee}. These proofs provide strong analytical
support for the notion of universality of the QHT.  This model of
antiferromagnetically interacting superspins has been shown to be critical
\cite{tsai}.  Numerically, the critical exponent $\nu$ was obtained from a
finite length scaling of superspin chains to be $\nu = 2.38 \pm .4$
\cite{kondev}. 

So far, no analytical information could be directly obtained on the critical
parameters, the localization exponent $\nu$ and the critical value $\Lambda_c$
of the scaling function. However, building on this model of a superspin chain,
supersymmetric conformal field theories have been suggested, which ultimately
should yield the critical parameters of the QHT 
\cite{conform1,conform2}.  

Restricting this theory to quasi-1D, by assuming a finite width $L_y$ of the
quantum Hall bar, of the order of the unitary noncritical localisation length
$\xi_{2D unit} = l \exp ( \pi^2 \sigma_{xx}^2)$, which serves as the
ultraviolet cutoff of the conformal field theory, one finds that the critical
value of the scaling function $\Lambda_c=1.2$ (the ratio of the localisation
length in a Quantum Hall wire and its finite width $L_y$, when the energy is
in the center of the Landau band, see Section 3.1) is fixed by the eigenvalues
of the Laplace-Beltrami operator of this supersymmetric conformal field theory
\cite{conform2,metzler}. This is a characteristic invariant of this theory,
arising from the conformal symmetry (just as the quantisation of angular
momentum arises from the rotational symmetry of a Hamiltonian).  Furthermore,
based on the properties of this constrained class of supersymmetric conformal
field theories, it has been predicted that the distribution function of local
wave function amplitudes is very broadly, namely log-normally, distributed.
This prediction has recently been confirmed by high-accuracy numerical
calculations \cite{parabol}.

The quest to derive the critical exponent of the localisation length at the
QHT from a critical supersymmetric theory has thus recently gained much
progress, but is still not yet complete .

\section{The mesoscopic quantum Hall transition}
Until here, the localization-delocalization transition in infinite systems
with periodic boundary conditions in one direction has been considered. Using
this model, it has been shown that disorder removes the degeneracy of the
Landau bands and introduces delocalized states only at one energy in each
band.

If Dirichlet boundary conditions are used the degeneracy of the Landau levels
is removed even without disorder due to the confining potential. The
eigenenergies form bands $\epsilon_{n}(X)$ which start at the positions of the
Landau levels and $\epsilon_{n}(X)\propto X^{2}$ for $X\gg L$. In the latter
energy region, the wave functions are 1D objects. They are delocalized along
the edges of the system. They have chiral character such that the
corresponding probability currents propagate in opposite directions at
opposite edges. In the perpendicular direction, the edge wave functions are
sharply localized with a distance $\ll \ell_{B}$. It has been shown that the
quantization of the Hall conductance can be understood in terms of these edge
states \cite{h82}. Commonly, it is argued that disorder does not influence
the edge states since forward and backward scattering are spatially
separated.

However, numerical data indicate that this is not always true. For short-range
disorder, edge and bulk-localized states can mix \cite{ook92}. Then, several
questions can be asked. For instance, can the extended wave functions near the
centers of the Landau bands coexist with the edge states in the presence of
disorder? If they don't, what are the localization properties of the mixed
states? Where do the 1D delocalized edge states go if the confining potential
is continously depleted?

In order to study such questions, quantum Hall wires with a finite width have
to be considered.  This has been done previously with the emphasis on
conductance fluctuations \cite{tetal89,a94}, edge state mixing
\cite{wetal91,ks93,rs95,oo88,ook89,a90,mk92} and the breakdown of the QHE
\cite{vetal87}. It has been suggested that edge wave functions might become
localized if in the presence of white noise disorder edge and bulk states mix
\cite{oo88,ook89,mk92}.

\begin{figure}[htbp]
   \includegraphics[width=7.5cm]{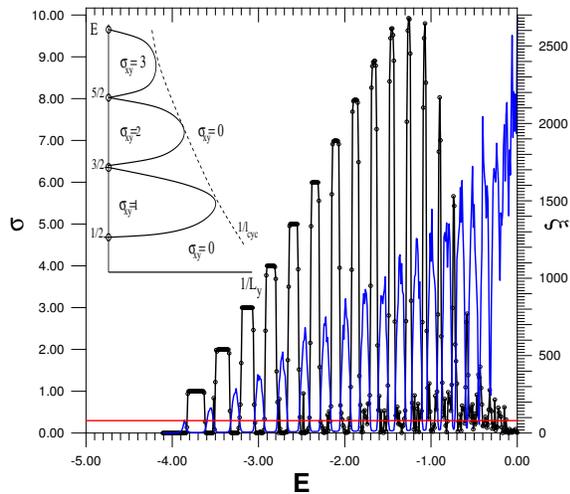}    
\caption[]{Black curve (left scale): two-terminal conductance $\sigma$ of a 
  quantum Hall wire described by the Peierls tight binding model (lattice
  constant $a$) with Dirichlet boundary conditions (width $L_y=80\,a$, length
  $L=5000\,a$, disorder $W=0.8\,V$, magnetic field 0.025 flux quanta per unit
  cell) in units of $e^2/h$ as a function of the energy $E$ (units $V$). Blue
  curve (right scale): localization length $\xi(E)$ calculated with the
  transfer matrix method for a quantum wire with periodic boundary conditions
  ($L\leq 100000\,a$); horizontal red line: $L_{y}=80\,a$. Inset: schematic
  phase diagram of the quantum Hall wire with $L\gg L_{y}$. Full lines:
  boundaries where $\sigma_{xy}$ jumps from $me^{2}/h$ to $0$ ($m$ integer,
  $L_{y}=\ell_{\rm cycl}$, dotted).}
  \label{fig:1storder}
\end{figure}

That this is indeed the case has been shown recently analytically and
numerically \cite{kko03}. In particular, it has been demonstrated that for
$T=0$ the two-terminal conductance and the Hall conductance of a quantum wire
in a magnetic field exhibit discontinuous transitions between the integer
plateau values and zero, for uncorrelated disorder and hard wall confinement,
as shown in Fig.~\ref{fig:1storder}. This is shown to be caused by
chiral-metal to insulator transitions of the quasi-1D edge states, driven by a
crossover from 2D to 1D localization of the bulk states. These metal-insulator
transitions resemble first-order phase transitions in the sense that the
localization length abruptly jumps between exponentially large and finite
values. In the thermodynamic limit {\em $\,$fixing the aspect ratio
  $c=L/L_{y}$ when sending $L\to\infty$}, and only then $c\to\infty$, the two
termional conductance jumps between exactly integer values and zero
conductance.  The transitions occur at energies where the localization length
of the bulk states is equal to the geometrical wire width, and $m$ edge states
can mix. When this happens, the electrons are free to diffuse between the wire
boundaries but become Anderson localized along the wire.

\section{Conclusion}
In conclusion, we have reviewed the standard models of localization in the
quantum Hall regime. We have provided an overview of the numerical results for
the critical exponent of the localization length. The fact that these were
obtained by using completely different models including white noise and long
range correlated disorder stongly suggests that the QHT is a universal quantum
phase transition. This also suggests that as long as interactions can be
treated on a mean field level the critical exponent is not changed. Many
experimental data indicate that the QHE is a manisfestation of this quantum
phenomenon. For quasi-1D quantum Hall systems, we have predicted that the
mixing of bulk and edge states near the centers of the Landau levels leads to
a crossover between 2D and 1D localization with drastic consequences for the
conductance: the two-terminal conductance drops sharply from the plateau value
to zero conductance at the energies where the bulk localization length is
equal to the geometrical width of the system. This {\em mesoscopic
quantum-Hall-insulator transition} should be observable in quantum wires with
short range disorder.

\section*{Acknowledgement} 
The authors would like to thank Prof. S. M. Nishigaki and PD Dr. B. Huckestein
for useful discussions. We thank Bodo Huckestein for providing the original of
Fig. \ref{kvkfig:2}. The work was supported by the Deutsche
Forschungsgemeinschaft via the Priority Programme ''Quanten-Hall-Systeme''
(Project Kr627/10) and by the EU via grants FMRX-CT98-0180 and
HPRN-CT2000-0144.
 The authors thank the Max Planck Institute for Physics of Complex Systems
 in Dresden for the hospitality during the completion of this article.


\begin{thebibliography}{9}
%
\bibitem{aalr79} E. Abrahams, P.W. Anderson, D. C. Licciardello, and T.V.
  Ramakrishnan, Phys. Rev. Lett. {\bf 42}, 673 (1979).
%
\bibitem{mk93} B. Kramer, A. MacKinnon, Rep. Progr. Phys. {\bf 56}, 1469
   (1993).
%
\bibitem{kvk80} K. von Klitzing, G. Dorda, M. Pepper, Phys. Rev. Lett. {\bf
    45}, 494 (1980).
%
\bibitem{kravchenko95}S. V. Kravchenko {\it et al.}, Phys. Rev. B{\bf 51},
  7038 (1995).
%
\bibitem{aa81} H. Aoki, T. Ando, Sol. St. Commun. {\bf 18}, 1079 (1981).
%
\bibitem{p81} R. E. Prange, Phys. Rev. B {\bf 23}, 4802 (1981). 
%
\bibitem{l81} R. B. Laughlin, Phys. Rev. B {\bf 23}, 5631 (1981).
%
\bibitem{h82} B. I. Halperin, Phys. Rev. B {\bf 25}, 2185 (1982).
%
\bibitem{wetal88} H. P. Wei, D. C. Tsui, M. A. Paalanen, and A. M. M.
  Pruisken, Phys. Rev. Lett. {\bf 61}, 1294 (1988).
%
\bibitem{ketal91} S. Koch, R. J. Haug, K. v. Klitzing, and K. Ploog, Phys.
  Rev. Lett. {\bf 67}, 883 (1991).
%
\bibitem{eetal93} L. W. Engel, et al., D. Shahar, C. Kurdak, and D. C. Tsui,
  Phys. Rev. Lett. {\bf 71}, 2638 (1993).
%
\bibitem{ketal0y} F. Kuchar, R. Maisels, T. Brandes, B. Kramer,
  Europhys. Lett. {\bf vv}, ppp (199y) 
%
\bibitem{f98} M. Furlan, Phys. Rev. B {\bf 57}, 14818 (1998).
%
\bibitem{hetal02} F. Hohls, U. Zeitler, and R. J. Haug, Phys. Rev. Lett. {\bf
    88}, 036802 (2002); F. Hohls et al., Phys. Rev. Lett. {\bf 89}, 276801
  (2002).
%
\bibitem{w83} F. Wegner, Z. Phys. B{\bf 51}, 279 (1983).
%
\bibitem{bgi84} E. Br\'ezin, D. J. Gross, C. Itzykson, Nucl. Phys. B {\bf
    235[FS11]}, 24 (1984).
%
\bibitem{l65} I. M. Lifshitz, Sov. Phys. Usp. {\bf 7}, 549 (1965).
%
\bibitem{aa85} T. Ando, H. Aoki, Phys. Rev. Lett. {\bf 54},
  832 (1985).
%
\bibitem{hk89} B. Huckestein, B. Kramer, Sol. St. Commun. {\bf 71}, 445
   (1989).
%
\bibitem{jp00} M. Janssen, K. Pracz, Phys. Rev. B {\bf 61}, 6278 (2000).
%
\bibitem{hk90} B. Huckestein, B. Kramer, Phys. Rev. Lett. {\bf 64}, 1437
   (1990).
%
\bibitem{m80} A. MacKinnon, J. Phys. C: Sol. St. Phys. {\bf 13}, L1031
   (1980).
%
\bibitem{mk81} A. MacKinnon, B. Kramer, Phys. Rev. Lett. {\bf 47}, 1546
   (1981).
%
\bibitem{mk83} A. MacKinnon, B. Kramer, Z. Phys. B {\bf 53}, 1 (1983).
%
\bibitem{km93} B. Kramer, A. MacKinnon, Rep. Progr. Phys. {\bf 56}, 1469
  (1993).
%
\bibitem{h95} B. Huckestein, Rev. Mod. Phys. {\bf 67}, 357 (1995).
%
\bibitem{aso02} Y. Asada,  K. Slevin, T. Ohtsuki, Phys. Rev. Lett {\bf 89},
256601 (2002).
%
\bibitem{ms88} G. V. Mil'nikov, I. M. Sokolov, Pis'ma Zh. Eksp. Teor. Fiz.
  {\bf 48}, 494 [JETP Lett. {\bf 48}, 536 (1988)].
%
\bibitem{h92} B. Huckestein, Europhys. Lett. {\bf 20}, 451 (1992).
%
\bibitem{p33} R. Peierls, Z. Phys. {\bf 80}, 763 (1933).
%
\bibitem{l51} J. M. Luttinger, Phys. Rev. B {\bf 84}, 814 (1951).
%
\bibitem{h76} D. R. Hofstadter, Phys. Rev. B {\bf 14}, 2239 (1976).
%
\bibitem{smk84} L. Schweitzer, B. Kramer, A. MacKinnon, J. Phys. C {\bf 17},
  4111 (1984).
%
\bibitem{mk85} A. MacKinnon, B. Kramer, Z. Phys. B {\bf vv}, ppp (1985).
\bibitem{t76} M. Tsukada, J. Phys. Soc. Japan {\bf 41}, 1466 (1976).
%
\bibitem{i82} S. V. Iordanski, Sol. St. Commun. {\bf 43}, 1 (1982).
%
\bibitem{kl82} R. F. Kazarinov, S. Luryi, Phys. Rev. B {\bf 25}, 7626 (1982).
%
\bibitem{o82a} Y. Ono, in:{\em Anderson Localization}, Springer Ser. Sol.
  State Sciences {\bf 39} (ed. by Y. Nagaoka, H. Fukuyama) p. 207 (Springer
  Verlag, Berlin 1982).
%
\bibitem{pj82} R. E. Prange, R. Joynt, Phys. Rev. B {\bf 25}, 2943 (1982).
%
\bibitem{t83}S. A. Trugman, Phys. Rev. B {\bf 27}, 7539 (1983).
%
\bibitem{s79} D. Stauffer, Phys. Rep. {\bf 54}, 2 (1979).
%
\bibitem{n79} M. P. M. den Nijs, J. Phys. A {\bf 12}, 1857 (1979).
%
\bibitem{be81} J. L. Black, V. J. Emery, Phys. Rev. B {\bf 23}, 429 (1981).
%
\bibitem{cc88} J. T. Chalker, P. D. Coddington, J. Phys. C{\bf 21}, 2665
  (1988).
%
\bibitem{f88} H. A. Fertig, Phys. Rev. B {\bf 38}, 996 (1988).
%
\bibitem{gr97} A. G. Galstyan, M. E. Raikh, Phys. Rev. B {\bf 56}, 1422
  (1997).
%
\bibitem{ajs97} D. P. Arovas, M. Janssen, B. Shapiro, Phys. Rev. B {\bf 56},
  4751 (1997).
%
\bibitem{jmmw98} M. Janssen, R. Merkt, J. Meyer, A. Weymer, Physica B {\bf
    256-258}, 65 (1998).
%
\bibitem{cr03} P. Cain, R. A. R\"omer, M. Schreiber, M. E. Raikh, Phys. Rev. B
  {\bf 64}, 235326 (2001); R. A. R\"omer, P. Cain, Adv. Sol. St. Phys. {\bf
    43}, 235 (Springer Verlag, Berlin 2003).
%
\bibitem{hc96} C. M. Ho, J. T. Chalker, Phys. Rev. B {\bf 54}, 8708 (1996). 
%
\bibitem{km95} R. Klesse, M. Metzler, Europhys. Lett. {\bf 32}, 229 (1995).
%
\bibitem{ff85} M. P. A. Fisher, E. Fradkin, Nucl. Phys. B {\bf 251[FS13]}, 457
  (1985).
%
\bibitem{f86} E. Fradkin, Phys. Rev. B {\bf 33}, 3257 (1986); ibid. 3263. 
%
\bibitem{lfsg94} A. W. Ludwig, M. P. A. Fisher, R. Shankar, G. Grinstein,
  Phys. Rev. B {\bf 50}, 7526 (1994). 
%
\bibitem{l94} D.-H. Lee, Phys. Rev. B {\bf 50}, 10788 (1994).
%
\bibitem{z95} K. Ziegler, Europhys. Lett. {\bf 28}, 549 (1995).
%
\bibitem{ac79} Y. Aharonov, A. Casher, Phys. Rev. A {\bf 19}, 2461 (1979).
%
\bibitem{ntw94} A. A. Nersesyan, A. M. Tsvelik, F. Wegner, Phys. Rev. Lett.
  {\bf 72}, 2628 (1994); NUcl. PHys. B {\bf 438}, 561 (1995).
%
\bibitem{b95} D. Bernard, Nucl. Phys. B {\bf 441}, 471 (1995).
%
\bibitem{ckt96} J.-S. Caux, I. I. Kogan, A. M. Tsvelik, Nucl. Phys. B {\bf
    466}, 444 (1996).
%
\bibitem{kmt96} I. I. Kogan, C. Mudry, A. M. Tsvelik, Phys. Rev. Lett. {\bf
    77}, 707 (1996).
%
\bibitem{o82} Y. Ono, J. Phys. Soc. Japan  {\bf 51}, 2055 (1982). 
%
\bibitem{m90} B. Miek, Europhys. Lett. {\bf 13}, 453 (1990).
%
\bibitem{hb92} Y. Huo, R. Bhatt, Phys. Rev. Lett. {\bf 68}, 1375 (1992).
%
\bibitem{hamg95} C. B. Hanna, D. P. Arovas, K. Mullen, S. M. Girvin,
  Phys. Rev. B {\bf 52}, 5221 (1995).
%
\bibitem{sm96} E. S. S{\o}rensen, A. H. MacDonald, Phys. Rev. B {\bf 54},
  10675 (1996).
%
\bibitem{lwk93} D.-H. Lee, Z. Wang, S. Kivelson, Phys. Rev. Lett. {\bf 70},
  4130 (1993).
%
\bibitem{b83} M. N. Barber, in: {\em Phase Transitions and Critical Phenomena},
    Vol {\bf 8}, ed. by C. Domb, J. Lebowitz (Academic London 1983), p. 146.
%
\bibitem{wtpp88} H. P. Wei, D. C. Tsui, M. A. Paalanen, A. M. M. Pruisken,
  Phys. Rev. Lett. {\bf 61}, 1294 (1988). 
\bibitem{pruiskentop} H. Levine, S. B. Libby, and A. M. M. Pruisken,
Phys. Rev. Lett.
 {\bf 51} , 1915 (1983);
    Nucl.  Phys.  B  {\bf 240}  30; 49; 71  (1984). 
%
\bibitem{ef}
K. B. Efetov, 
{\it Supersymmetry in Disorder and Chaos} Cambridge University Press, 
Cambridge (1997).
%
\bibitem{weiden}  H. A. Weidenm\" uller,
 Nucl. Phys. B {\bf 290}, 87 (1987);
H. A. Weidenmueller and  M. R. Zirnbauer,
Nuclear Physics B { \bf 305}, 339 (1988). 
%
\bibitem{khm}  D. E. Khmel'nitskii,
JETP Lett. {\bf 38}, 552 (1983). 
%
\bibitem{affleck} I. Affleck, Nucl. Phys. {\bf B 265}, 409 (1986). 
%
\bibitem{tsvelik} S. Kettemann and A. Tsvelik, Phys. Rev. Lett. {\bf 82},
3689 (1999).   
%
\bibitem{marikhin} K. B. Efetov, and V. G. Marikhin, Phys. Rev. {\bf B 40},
12126 (1989).  
%
\bibitem{zirnchain} M. R. Zirnbauer, 
 Ann. d. Phys. {\bf 3}, 513 (1994).
%
\bibitem{lee}  D.H. Lee, Phys. Rev.  {\bf B 50}, 10 788 (1994);
     J. Kondev and J.B. Marston, Nucl. Phys. {\bf B 497}, 639 (1997); 
     M.R. Zirnbauer, J. Math. Phys. {\bf 38}, 2007 (1997).
%
\bibitem{tsai} J. B. Marston and Shan-Wen Tsai, 
           Phys. Rev. Lett. {\bf 82}, 4906 (1999).
%
\bibitem{kondev} J. Kondev and J. B. Marston, Nuclear Physics  {\bf B  497},
  639 (1997). 
%
\bibitem{conform1} M. Zirnbauer, hep-th/9905054.
\bibitem{conform2}
  M.J. Bhaseen, I. I. Kogan, O. A. Soloviev, N. Taniguchi and A. M. Tsvelik,
 Nucl. Phys. {\bf B 580}, 688 (2000).
%
\bibitem{klesse}       R. Klesse, M. R. Zirnbauer,       Phys. Rev. Lett. {\bf 86}, 2094 (2001).
%
\bibitem{metzler}  M. Janssen, M. Metzler, M. R. Zirnbauer
 Phys. Rev. {\bf B 59}  15836 (1999).
%
\bibitem{parabol}  F. Evers, A. Mildenberger, and  A.D. Mirlin,
  Phys. Rev. {\bf B 64}, 241303(R) (2001).
%
\bibitem{ook92} Y. Ono, T. Ohtsuki, B. Kramer, in: {\em High Magnetic Field in
    Semiconductor Physics}, ed. by G. Landwehr, Springer Ser. Sol. St. Sci.
  {\bf 101}, p. 60 (Springer, Berlin 1992).
%
\bibitem{tetal89} G. Timp, A. M. Chang, P. Mankiewich, R. Behringer, J. E.
  Cunningham, T. Y. Chang, R. E. Howard, Phys. Rev. Lett. {\bf 59}, 732
  (1987); G. Timp, R. Behringer, J. E. Cunningham, R. E. Howard, 
    Phys. Rev. Lett. {\bf 63}, 2268 (1989).
%
\bibitem{a94} T. Ando, Phys. Rev. B {\bf 49}, 4679 (1994). 
%
\bibitem{wetal91} B. J. van Wees, L. P. Kouwenhoven, E. M. M. Willems, C. J.
  P. M. Harmans, J.  E. Mooij, H. van Houten, C. W. J. Beenakker, J. G.
  Williamson, C. T. Foxon, Phys. Rev. B {\bf 43}, 12431 (1991).
%
\bibitem{ks93} B. I. Shklovskii, Pis'ma Zh. Eksp. Teor. Fiz. {\bf 36}, 43
  (1982) [ JETP Lett. {\bf 36}, 51 (1982)]; A. V. Khaetskii and B. I.
  Shklovskii, Zh. Eksp. Teor. Fiz. {\bf 85}, 721 (1983) [Sov. Phys. JETP {\bf
    58}, 421 (1983)].
%
\bibitem{rs95} M. E. Raikh, T. V. Shahbazyan, Phys. Rev. B {\bf 51},
  9682 (1995).  
%
\bibitem{oo88} T. Ohtsuki, Y. Ono, Sol. St. Commun.  {\bf 65}, 403 (1988);
  ibid. {\bf 68}, 787 (1988).
%
\bibitem{ook89} Y. Ono, T. Ohtsuki, B.  Kramer, J. Phys. Soc. Jpn. {\bf 58},
  1705 (1989); T. Ohtsuki, Y. Ono, J. Phys. Soc. Jpn. {\bf 58}, 956 (1989).
%
\bibitem{a90} T. Ando, Phys. Rev. B {\bf 42}, 5626 (1990).
%
\bibitem{mk92} R. G. Mani, K. v. Klitzing, Phys. Rev. B {\bf 46}, 9877
  (1992); R. G. Mani, K. von Klitzing, K. Ploog, Phys. Rev. B {\bf 51},
  2584 (1995).
%
\bibitem{vetal87} P. G. N. d. Vegvar, A. M. Chang, G. Timp, P. M. Mankiewich,
  J. E. Cunningham, R. Behringer, R. E. Howard, Phys. Rev. B {\bf 36},
  9366 (1987).
%
\bibitem{kko03} S. Kettemann, B. Kramer, T. Ohtsuki, unpublished (2003).

\end{thebibliography}
\end{document}